\documentstyle[12pt]{article}
\renewcommand{\baselinestretch}{2.0}
\def \be{\begin{equation}}
\def \ee{\end{equation}}
\def \bea{\begin{eqnarray}}
\def \eea{\end{eqnarray}}
\begin{document}
\hskip 9 cm IPM-96-150
\begin{center}
{\Large{\bf{ A Logarithmic Conformal Field Theory Solution For Two Dimensional
Magnetohydrodynamics in Presence of The Alf'ven Effect}}}
\vskip 1cm
{\large{ M. R. Rahimi Tabar$^{1,2}$ \\and \\S. Rouhani$^{1,3}$ }}
\vskip 1cm
\small{{\it{1)Institue for Studies in Theoretical Physics and 
Mathematics\\ Tehran P.O.Box: 19395-5746, Iran.\\2)Dept. of Physics, Iran
University of Science and Technology,\\ Narmak, Tehran 16844, Iran.
\\3)Department of Physics, Sharif University of Technology\\
Tehran P.O.Box:11365-9161, Iran. }}}
\end{center}
\vskip 1.5cm
\begin{abstract}
When Alf`ven effect is peresent in magnetohydrodynamics
 one is naturally lead to consider conformal field theories, which
have logarithmic terms in their correlation functions. We discuss the
implications of such logarithmic terms and 
find a unique conformal field theory with centeral charge 
$c=-\frac{209}{7}$, within the border of the minimal series, which
satisfies all the constraints.
The energy espectrum is found to be \newline $E(k)\sim k^{-\frac{13}{7}} \log{k}$.
\end{abstract}
\newpage
\renewcommand{\baselinestretch}{2.0}
\topmargin -20mm
\oddsidemargin -7mm
\textwidth 165mm
\textheight 230mm
\def \be{\begin{equation}}
\def \ee{\end{equation}}
\def \bea{\begin{eqnarray}}
\def \eea{\end{eqnarray}}
{\bf{1 - Introduction}}

There has been some work on modelling turbulence in two dimensional fluids
by conformal field theory (CFT) [1-3].
Ferretti et al. [4] have generalized Polyakov's method [1] to the case of two
dimensional magnetohydrodynamics (2D - MHD).
We have argued that the existence of a critical dynamical index is
equivalent to the Alf'ven effect [5] i.e. the equipartition of energy between
velocity and magnetic modes [6].
The Alf'ven effect, reduces the number of candidate conformal field theories,
but also it implies that the velocity stream  function $\phi$ and the
magnetic flux function $\psi$ should have similar scaling dimensions.
This naturally leads to logarithmic conformal field theories, Gurarie [7] 
has argued that such theories do exist albeit the need to extend some of the 
definitions of CFT to include logarithmic operators [8].  

In such conformal field theories, it has been shown [7] that the correlator
of two fields, has a logarithmic singularity.
\be
<\psi(r) \psi(r^{'})>\sim {|r-r^{'}|}^{-2 h_\psi} \log|r-r^{'}| + \ldots
\ee
Examples of such theories have been studied by many authors [9-18], 
a particularly 
interesting application is in the field of disordered systems [17,18].
In this paper we shall present a CFT with central charge $c=-\frac{209}{7}$, 
which is a member of the series $c_{p,1}$ [8]. This is the only CFT found 
so far which satisfies all the constraints of 2D-MHD, including the Alf'ven
effect. In this CFT logarithmic correlators appear and in particular the 
energy spectrum is as follows:
\be
E(k)\sim k^{-\frac{13}{7}} \log{k}
\ee
A logarithmic dependence was observed in computer solutions by Borue[19].                                                
\\
This paper is organised as follows; in section two we give a very brief
summary of magnetohydrodynamics and the Alf'ven effect. In section 3 
we discuss the
implication of the logarithmic divergence and candidate CFT models are given
in section 4 .

\vskip 1.5cm
{\bf{2 - The Alf'ven effect and conformal field theory.}}

The incompressible two dimensional magnetohydrodynamic (2D - MHD) system has
two independent dynamical variables, the velocity stream function
$\phi$ and the
magnetic flux function $\psi$. These obey the pair of equations [20],
\be
{{\partial \omega} \over{\partial t}} = -e_{\alpha \beta} \partial_\alpha \phi
\partial_\beta \omega + e_{\alpha \beta} \partial_\alpha \psi \partial_\beta
J + \mu {\bigtriangledown}^2 \omega
\ee
\be
{{\partial \psi}\over{\partial t}} = -e_{\alpha \beta} \partial_{\alpha}\phi
\partial_\beta \psi + \eta J
\ee
where the vorticity $\omega = \bigtriangledown^2 \phi $ and the current 
$J =
\bigtriangledown^2 \psi$. The two quantitiesy $\mu$ and $\eta$ are the
viscosity
and molecular resistivity, respectively. The velocity and magnetic fields are
given in terms of $\phi$ and $\psi$ :
\be
V_\alpha = e_{\alpha \beta} \partial_\beta \phi
\ee
\be
B_\alpha = e_{\alpha \beta} \partial_\beta \psi
\ee
and $e_{\alpha \beta}$ is the totally antisymmetric tensor, with $e_{12} = 1$.
Chandrasekhar [5] has shown that the Alf'ven effect or the equipartition of
energy between velocity and magnetic modes requires $V^2_k = \alpha B^2_k$, 
with
$\alpha$ of order unity.
In fact he finds $\alpha = 1.62647$ for 2D - MHD. We [6] have argued
that
the existence of a critical dynamical index for 2D - MHD, implies the Alf'ven
effect and if the conformal model holds, this implies the equality of scaling
dimensions of $\phi$ and $\psi$ :
\be
h_\phi = h_\psi
\ee
Here the criteria
of Gurarie [7] are satisfied and these two fields are
logarithmically correlated. 
According to Gurarie [7],
the operator product expansion of two fields $A$ and $B$,
which have two fields
$\phi$ and $\psi$ of equal dimension in their fusion rule, has a 
logarithmic
term:
\be
A(z) B(0) = z^{h_\phi - h_A - h_B}\{\psi(0) + \ldots +\log z(\phi(0) + \ldots )
\}
\ee
to see this it is sufficient to look at four point function :
\be
<A(z_1) B(z_2) A(z_3) B(z_4)> \sim {1\over {(z_1-z_3)}^{h_A}} {1 \over
{(z_2 - z_4)}^{h_B}} {1 \over {[x(1-x)]^{h_A + h_B - h_\phi}}}
F(x)
\ee
Where the cross ratio $x$ is given by :
\be
x = {{(z_1 - z_2)(z_3 - z_4)}\over{(z_1 - z_3)(z_2 - z_4)}}
\ee
In degenerate 
models $F(x)$ satisfies a second order linear differential equation.

The hypergeometric equation governing the
correlator of two fields in whose OPE, two other fields $\psi$ and $\phi$ 
with conformal dimension $h_{\psi}$ and $h_{\psi}+\epsilon$ appear,
admits two solutions [10]:
\be
 _{2}F_{1} (a, b, c, x) 
\ee
\be
x^{\epsilon}  _{2}F_{1}(a+\epsilon , b+\epsilon, c+2\epsilon, x)
\ee
where a, b, and c are sums of conformal dimensions. Clearly in the limit of 
$\epsilon \rightarrow 0$ these two solutions coincide. Another independent
solution exists, it involves logarithms and can be
generated by standard methods.
Therefore in which case two
independent solutions can be constructed according to :
\be
\sum b_n x^n + \log x \sum a_n x^n
\ee
Now consistency of equation (12) and (8) requires :
\be
<A(z_1) B(z_2) \psi(z_3)> = <A(z_1) B(z_2) \phi(z_3)>\{ \log{(z_1-z_2) \over
{(z_1-z_3)(z_2-z_3)}} + \lambda \}
\ee
\be
<\psi(z) \psi(0)> \sim {1 \over {z^{2h_\psi}}} [\log z +\lambda^{'}]
\ee
\be
<\psi(z) \phi(0)> \sim {1 \over {z^{2 h_\phi}}}
\ee
where $\lambda$ and $\lambda^{'}$ are constants.  \\
Note that the correlators of this theory are annihilated by the set  
$(L_{-1}, L_0^2, L_- L_0 , L_+L_0 )$, thus we may solve a differential 
equation for$ <\psi(z_1) \psi(z_2)>$ , which leads to logarithmic 
singularities [18,21].
This is compatible with the findings in [7] that this type of 
operators together with ordinary primary operators form the basis of 
the Jordan cell for the operator $L_0$. This fact allows us to find 
higher-order correlation functions for the operator $\psi$ [21].
\\
Thus a candidate CFT has to be logarithmic due to condition imposed by eq.(7) 
, the fields $\phi$ and $\psi$ must have negative conformal dimensions and also 
satisfy the cascade condition. \\ 
Consider the fusion of two fields $\phi$ and $\psi$ :
\be
  \phi \times  \psi = \chi + ....
\ee
Such that $\chi$ is the field with minimum conformal dimension, on the right 
hand side. Then the magnetic potential cascade implies[6]:
\be
\Delta_{\phi} + \Delta_{\chi} = -2
\ee
The satisfaction of this set of constraints is considered in section 4, 
but first we must reconsider the problem of infra red divergence.  
\vskip 1.5cm
{\bf{3- The Infrared problem and The Energy Spectrum:}}

The presence of logarithmic terms requies a reconsideration of the
infrared problem. The $k$-representation of the correlation is;
\be
<\psi(k) \psi(-k)> = |k|^{-2-2|h_\phi|} [C_1 + \log k]
\ee
which is divergent in the limit of $k\rightarrow 0$ .
One can set some cut-off in the $k$-space to remove this divergence :
\bea
<\psi(x) \psi(0)> &=& \int^\infty_{k>{1\over R}} k^{-2-2|h_\phi|}
[C + \log k] e^{ik\cdot x} d^2 k \cr
&\sim & R^{2|h_\phi|} (\log R +C^{'}) - x^{2|h_\phi|}(C^{'} + \log X) 
+ \ldots
)
\eea
where $R$ is the large scale of the system.
It seems that it is natural to add some condensate term [1] in momentum space
to cancel the infrared divergence.
The energy spectrum for this type
of correlation, is
\be
E(k) \simeq  k^{-2|h_\phi|+1}(C +\log k)
\ee
This spectrum is compatible with the results of Ref. [19] where it has been
shown that, one loop correction to the energy spectrum gives a
logarithmic contribution to the energy spectrum.
\\
{\bf{4- Finding a Candidate Conformal Field Theory.}}\\
A possible candidate may exist within the $c_{p,1}$ series[8,21]. The central charge 
for this series is $c=13-6(p+p^{-1})$. This series is particular since it 
has $c_{eff}=1$. These CFT's posess $3p-1$ highest weight representations 
with conformal dimensions:
\be 
h_{p,s}= \frac{(p-s)^2 - (p-1)^2}{4p} \hskip .5cm,  1 \leq s \leq 3p-1
\ee
of these $2(p-1)$ have pair wise equal dimensions. Two fields $\phi_{s}$ and
$\phi_{s'}$ have equal and negative weights provided that 
$s+s'=2p$ , $ (s \neq 1, 2p-1)$ . Let us adopt such a pair as candidates for the 
fluxes $\phi$ and $\psi$. Then the fusion rule gives: 
\be 
 \phi_{s} \times  \phi_{s'} = \phi_{2p-1} + \cdots +\phi_{1}
\ee
where the sum is only over the odd values .
The field on the right hand side of eq.(23) with the lowest dimension is a 
candidate for $\chi$. We then have the dimension of $\chi$ as:
\be
  \Delta_\chi= \left\{ \begin{array}{ll} {-k^2 \over 2k+1} & p=2k+1 \\
  -\frac{1}{2}(k-1) & p=2k \end{array} \right. 
\ee
We can now look for the candidate values of $s$ and  $p$ such that eq.(18) 
is satisfied. The only solution is given by: 
\be
  p=7 \hskip 1 cm,  \Delta_\phi =\Delta_\psi =-\frac{5}{7} \hskip 1cm, 
  \Delta_\chi =-\frac{9}{7}
\ee
Note that for this solution $\chi$ is the field with minimum conformal 
dimension in this CFT. Input these values into the formula for the energy 
spectrum to get:  
\be
   E(k)\sim k^{-\frac{13}{7}} \log{k}
\ee

In summing up we observe that the imposition of the Alf'ven effect greatly 
narrows the choice for candidate CFT's. From an infinte number of candidates 
with $h_{\psi} \ne h_{\phi}$ , we end up with just one candidate with 
$h_{\psi} = h_{\phi}$.


\newpage

\begin{thebibliography}{99}
\bibitem{1} A. M. Polyakov Nucl. Phys. {\bf{B 396}}, 397 (1993)
\bibitem{2}  D. A. lowe, Mod. Phys. lett. {\bf{A8}}, 923 (1993)
\bibitem{3}  B. K. Chung, S. Nam, Q. H. park and H. J. Shin, Phys. lett. 
{\bf{B 309}},
58(1993), Chia-Chu Chen, Mod. Phy. lett. {\bf{A9}}, 123(1994).
\bibitem{4}  G. Ferretti, Z. Yang, Europhy. lett. {\bf{22}} (9), 639 (1993)
\bibitem{5} Chandrasekhar, S. Annals of Physics, {\bf{ 2}} 615, (1957)
\bibitem{6}  M. R. Rahimi Tabar and S. Rouhani Annals of Physics {\bf{ 246}}, 446 (1996)
\bibitem{7} V. Gurarie, Nucl. Phys. {\bf{B410}}, 535 (1993)   
\bibitem{8} M.A.I. Flohr, "On Modular Invariant Partition Function of 
Conformal Field Theory with Logarithmic Operators", CSIC preprint, hep-th/9509166

\bibitem{9} O. Coceal, W. Sabra and S. Thomas " Conformal Solutions
of Duality Invariant 2D Magnetohydrodynamic
Turbulence" QMW - PH - 96 - 05.
\bibitem{10} M. R. Rahimi Tabar and S. Rouhani "The Alf'ven Effect and 
Conformal Field Theory", IPM-94-95, hep-th/9507166.  

\bibitem{11} H. Saleur, Nucl. Phys. \bf{B382}, 486 (1992),\bf{B382}, 532 (1992) 
\bibitem{12} L. Rozansky, H. Saleur Nucl. Phy. {\bf{B 376}}, 461 (1992)
\bibitem{13} G. Cardy, UCSB Preprint UCSBTH- 91- 56, 1991  
\bibitem{14} X. G. Wen, Y. S. Wu and Y. Hatsugai, Nucl. Phys.{\bf{B422}}[FS] 
,476 (1994)
\bibitem{15} J. Ellis and N. Mavromatos, "D-branes from Liouville Strings",
hep-th/9605046   
\bibitem{16} I.I. Kogan and N. Mavromatos, Oxford preprint-95-50P,hep-th/9512210   
\bibitem{17} J.S. Caux, I.I. Kogan and A.M. Tsvelik, "Logarithmic
Operators and Hidden Symmetry in Critical Disordered Models", Oxford 
Preprint OUTP-95-62S, hep-th/9511134
\bibitem{18} Z. Maassarani and D. Serban "Non-unitary Conformal Field Theory
and Logarithmic Operators for Disordered Systems", SPHT-T96/037, 
hep-th/9605062
\bibitem{19} V. Borue, Phys. Rev. Lett.{\bf{71}}, 3967 (1993) 

\bibitem{20} V. N. Tsytovich." Theory of Turbulent Plasma" Consultants Bureau,
New York (1977)

\bibitem{21} A. Shafiekhani and M. R. Rahimi Tabar,"Logarithmic Operators in 
Conformal Field Theory and The $W_{\infty}$-algebra"hep-th/9604007
\bibitem{22} H. G. Kausch Phys. Lett. {\bf B 259},448 (1991) 
\end{thebibliography}
\end{document}